# Improved Spectral Imaging Microscopy for Cultural Heritage through Oblique Illumination

Lindsay Oakley[1], Stephanie Zaleski[1], Billie Males[1], Ollie Cossairt[2], and Marc Walton[1*]
   1) Center for Scientific Studies in the Arts, Northwestern University
   2) Department of Computer Science, Northwestern University

*Marc Walton.
Email: marc.walton@northwestern.edu

**Abstract**

This work presents the development of a flexible microscopic chemical imaging platform for cultural heritage that utilizes wavelength-tunable oblique illumination from a point source to obtain per-pixel reflectance spectra in the VIS-NIR range. The microscope light source can be adjusted on two axes allowing for a hemisphere of possible illumination directions. The synthesis of multiple illumination angles allows for the calculation of surface normal vectors, similar to phase gradients, and axial optical sectioning. The extraction of spectral reflectance images with high spatial resolutions from these data is demonstrated through the analysis of a replica cross-section, created from known painting reference materials, as well as a sample extracted from a painting by Pablo Picasso entitled *La Miséreuse accroupie* (1902). These case studies show the rich microscale molecular information that may be obtained using this microscope and how the instrument overcomes challenges for spectral analysis commonly encountered on works of art with complex matrices composed of both inorganic minerals and organic lakes.

**Introduction**

The technical study of objects from art and archaeology often involves the removal of microsamples for chemical analysis and imaging, which provide fundamental microstructural information to explore how these objects were created and to determine how their constituent materials may deteriorate over time.(1, 2) Cross-sections of artworks often present a complicated heterogeneity of organic and inorganic components, and it is therefore beneficial to identify each component with spatial sensitivity. Analytical techniques such as scanning electron microscopy/energy dispersive x-ray spectroscopy (SEM/EDS) can provide element analysis at high resolution, but molecular information and the identity of organic constituents must be obtained from other analytical methods such as Raman or Fourier Transform Infrared (FTIR) micro-spectroscopy.(3–5) While Raman or FTIR instruments have been configured for high-resolution imaging,(6–9) these measurements can be time consuming and consequently within most cultural heritage laboratories these techniques are primarily used for point analysis rather than molecular mapping. Similarly, to achieve elemental images via EDS mapping routines requires exposure of a sample to an electron beam for prolonged periods of time necessary to achieve sufficient counting statistics. These conditions often cause radiative damage that precludes further imaging and analysis of the same sample location.(1, 10) As an alternative to these traditional workhorse techniques of analysis, here it is shown that microscale chemical maps can be produced using less damaging optical wavelengths through a spectral imaging microscope. We build on previous work in cultural heritage (11) towards highly sensitive spectral reflectance imaging at the microscale by producing an instrument that provides both traditional optical images of microscopic features, as well as material maps, with as little as possible radiative damage to the sample. We also demonstrate the use of computational techniques to obtain diffraction-limited chemical images of both organic and inorganic materials present in cross-sections in a single experiment.

The majority of optical microscopy experiments in cultural heritage are performed in reflection where the cross-section is prepared as a thick polished slab.(1) Because polished cross-sections are excellent specular reflectors, it is necessary to configure the sample illumination to suppress such reflections which carry none of the desired molecular information. While it is possible to avoid specularity using a transmission illumination geometry, this method requires thin sections prepared by microtomy. Although possible, the material properties of some samples can make them challenging to section and often the necessary cutting instrumentation is not readily available in art conservation laboratories.(12–14) Careful consideration of the limitations and advantages of specular suppression methods informed the design of the spectral microscope described here in order maximize signal and spatial resolution.

One method to eliminate specular reflections is through the use of cross-polarization filters. Specular reflections that retain the polarization state of the source light are efficiently rejected by a second filter oriented orthogonally to a source filter, while randomly polarized light rays associated with diffuse scattering are allowed.(15, 16) However, precise crossed alignment is required to achieve peak efficiency and maximize the number of photons collected for imaging.(17, 18) More importantly, neither the extinction nor transmission efficiency of most polarization filters is uniform across the visible range which is not ideal for spectral microscopy over wide wavelength ranges (*e.g.*, 400-1000 nm). Standard widefield polarized light microscopes also have a large depth of field which contributes to a large volume of forward scattered light (on the order of 10-100s of microns) and significantly increases spectral blur based on the absorption properties of the sample matrix. To our knowledge, these large interaction volumes inherent to axial reflection-geometries of a standard widefield microscope are the main reason spectral microscopy has not been adopted as a micron-resolution analytical technique in cultural heritage science.

As an alternative to bright field microscopy, dark field (DF) spectral microscopes have been demonstrated in the context of biological applications with excellent spatial resolution, contrast, and shallow light penetration depths, providing a promising avenue for the examination of cross-sections.(19–25) In DF microscopy, a hollow cone of light with a numerical aperture (NA) larger than the objective illuminates the sample. Usually a beam stop is placed in the illumination path to match the size of the back aperture to create these conditions. Progressively dimmer images are produced with increasing magnification as the diameter of the beam stop must increase with the NA of the objective. Such a light path can be problematic when trying to efficiently recover light reflected from a highly absorbing sample, such as pigments routinely encountered in objects from cultural heritage. Additionally, while the DF setup should prevent diffracted light (less than first order) from entering the objective,(26) certain orientations of microfacets due to surface roughness can still guide light into the objective and contaminate the image with specular reflections.

Recent innovations have achieved enhanced contrast and spatial resolution through tunable-angle oblique-illumination (OI) as an alternative to DF illumination.(26, 27) OI from a single point source can increase contrast and enhance details in the direction of incident illumination.(27) It has also been demonstrated that optimizing the angle of OI increases sensitivity based on the particle size and shape of the scattering element.(28, 29) Ma et al. has drawn on recent advances in oblique ptychographic illumination from an annular array of lights to improve the confocal gate of a standard brightfield microscope.(30) They demonstrated that individual point sources (LEDs) of monochromatic light retain partial coherence and thus improve the lateral resolution of the microscope. When multiple sub-images from each illumination direction were synthetically combined, the contrast of the in-focus elements was strongly enhanced while out-of-focus background was smeared, thus improving the microscope's overall axial sectioning capabilities. Using these principles, an OI spectral imaging microscope will allow for enhanced absorption contrast at the sample surface, thus overcoming out-of-plane contributions to spectral blurring that are encountered in standard widefield microscopes.

The spectral microscope we present here implements OI using mobile axes of illumination, as demonstrated by the analysis of paint cross sections of both known reference materials and a sample from an early 20$^{th}$ century painting by Pablo Picasso. These samples provide challenging case studies with complex heterogeneity at the microscale and imperfect surface features, which aids in highlighting the usefulness of OI in separating subsurface light/matter interactions from the image plane of interest. In addition, OI successfully captures spectral signatures of organic pigments that would have been otherwise missed by elemental analysis alone.

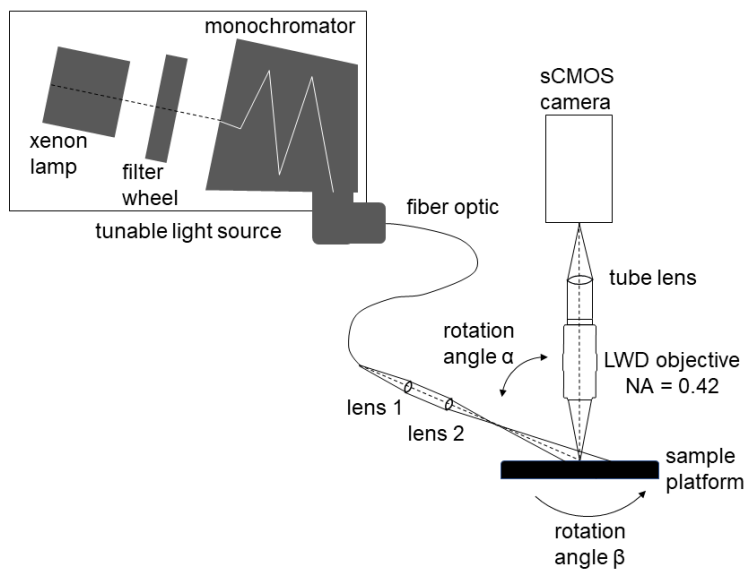

**Figure 1.** Schematic of the oblique illumination spectral microscope. The tunable illumination angle can be adjusted to change the angles of light in the polar direction as indicated by "rotation angle α" and the sample can be rotated in the azimuthal direction ("rotation angle β") to produce multiple views of the specimen with different lighting angles.

## Results and Discussion

*OI Microscope Design Approach*.

     A schematic of the microscope is shown in **Figure 1**, where the motorized movement of the illumination optics is indicated as "rotation angle α". This illumination arm is decoupled from the detection axis of the microscope in order to send incident light to the sample obliquely from adjustable polar angles. Illumination around the specimen was achieved by rotating the sample to multiple azimuthal angles ("rotation angle β", **Figure 1**) and taking an image at each position. Altogether, this multi-angle dome of illumination positions produces a large solid angle of light delivered to the sample in reflection and enables extraction of diffuse reflection measurements from multiple views of the specimen. We implement a tunable light source not only for the advantageous possibility to conduct excitation and emission studies to characterize specimen autofluorescence

or luminescence, but also because the reflected light is then collected directly by the camera without additional dispersive optical elements or specialized objectives.

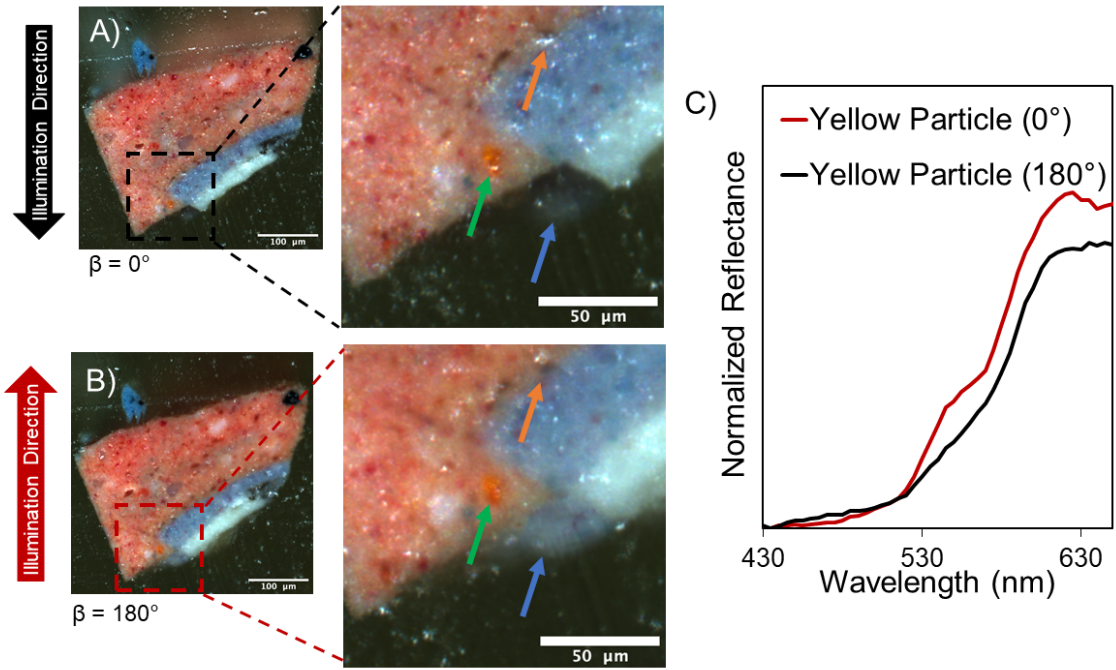

**Figure 2.** RGB images of a cross section from Picasso's *La Miséreuse accroupie* acquired at A) β = 0° and B) β = 180°. The arrows highlight distinct differences between images due to illumination direction. C) Reflectance spectra extracted from the yellow-orange particle at β = 0° (red trace) and β = 180° (black trace). See Movie S1 for multiple illumination angles.

The key idea behind the design of the microscope is that light reflected and scattered from sample's surface is dependent on the oblique angle of illumination. The effect of illumination angle on acquired spectral data is illustrated in **Figure 2** with a cross-section previously described by Pouyet et. al. (31) which was removed from the tacking edge of Pablo Picasso's 1902 painting entitled *La Miséreuse accroupie,* painted during his Blue Period. From this cross-section, spectral image stacks were collected, and two illumination angles were selected and collapsed into separate RGB images: **Figure 2A** and **2B** show the sample illuminated at β = 0° and 180°, respectively. Some significant differences between the 0° and 180° views are indicated by arrows in the enlarged insets: the orange arrow shows dissimilar interreflection highlights caused by surface concavities around a void, and the blue arrow indicates a feature below the embedding resin missing in the 0° image yet appearing at 180°. The green arrow points toward a yellow-orange particle from which the reflectance spectra, shown **Figure 2C**, were extracted. The spectrum at β = 0° has a shoulder at 535 nm which is greatly diminished at 180°. These observed changes in reflectance can be attributed to interreflections, subsurface scattering, and volumetric scattering that fluctuate with the illumination angle (see Movie S1). However, diffusely reflected light from the in-focus plane should remain constant at all angles in our observation geometry, provided that the sample is topographically flat and the surface normal vector is aligned with the optical axis.(32) Here, following a simple calibration procedure, we suggest that a minimum projection through all images can provide an upper bound on the estimated value of directly reflected diffuse light based on the premise that these other scattering mechanisms listed above only add to the intensity of a given pixel.(33) It corresponds that taking the difference of a maximum and minimum image provides an estimation of these effects.(34, 35) The capability to separate light scattered from different focal planes is of significance to the design of our microscope because this light is responsible for undesired spectral blurring and prevents the collection of reflectance spectra at the highest possible spatial resolutions.(36–39)

*Surface Shape Estimation and Min-Max Projections.*
Although careful sample preparation can produce samples with mirror-like finishes to the unaided eye, based on the resolution of our microscope, even techniques like microtomy or mechanical polishing with fine diamond slurries will leave cross sections with surface microfacets that can introduce angle dependent highlights and shadow artifacts. Light reflection from these facets can contribute to an over- or under-estimation of diffusely reflected light depending on whether it is a topographical peak or valley, respectively, in the minimum projection of the image stack. Likewise, dependent on the orientation of the surface to the detection axis, there will be a cosine falloff of illumination, following Lambert's law, that requires calibration to compensate. Such topography and sample orientation, however, can be computationally flattened by normalizing the images to the surface shape gradient at each angle of illumination.

Here we estimate the surface shape by using uncalibrated photometric stereo (PS) given by Equation 1:

$$I = \mathcal{N} \cdot \mathcal{L}, \qquad \mathcal{N} = kN \qquad (1)$$

In this parameterization, the albedo (diffuse color) $k$ is absorbed into the surface normal so $|\mathcal{N}| = k$. $I$ is the image intensity given by $I(x,y)$ under the illumination vector $\mathcal{L}$. $\mathcal{N}$ is recovered by solving Equation 1 via least squares given the known lighting directions.(40)

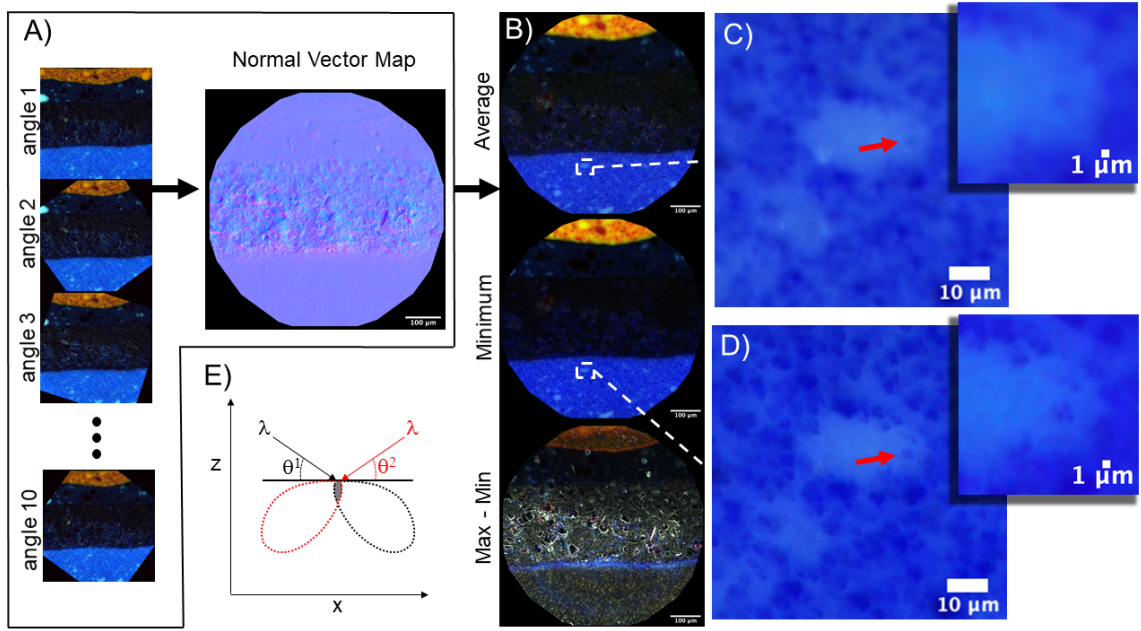

**Figure 3.** Spectral image processing workflow of a reference cross-section. A) Images were acquired in 10 angle increments from β = 0-360°, aligned, and used to calculate a surface normal vector map. Using this normal map, gradients were calculated for every illumination angle and then used to normalize the input images. B) From the surface shape corrected image set, an average projection (darkfield equivalent), minimum projection (focal plane diffuse scattering only), and max-min (extraneous additional scatter) images were created. C) Detail of the average image shows blur in the 1-10 μm range, but the minimum image displays pigment particles with resolvable features below 1 μm. E) The schematic diagram represents the overlap of two scattering paths. The minimum projection provides a measure of the overlap between the scattering regions (gray region).

**Figure 3A** shows a series of RGB images of a reference cross-section comprised of four layers with pigment compositions indicated in **Table S2**. From these images, a normal vector map was calculated which demonstrates that there is measurable surface texture in all four layers, even after polishing with slurries down to 0.5 µm. The most significant texture seems to be isolated to the second layer which contains a high concentration of the organic pigment, madder lake. Based on this observed morphology, the polishing procedure employed may have selectively dissolved the water-soluble organic components.

In addition to qualitative shape information provided by the normal map, surface gradients, which are the dot product of the surface formal vector and lighting direction ($\mathcal{N} \cdot \mathcal{L}$), were calculated for each illumination direction. These gradients were then used to normalize the input images prior to calculating the output average, minimum, and difference as presented in **Figure 3B**. More formally, the diffuse color $k$ is calculated for every illumination angle ($l$) and wavelength ($j$) following,

$$k_{lj} = I_{lj}(\mathcal{N}_{xyz} \cdot \mathcal{L}_l)^{-1} \qquad (2)$$

where $\mathcal{N}_{xyz}$ is the normal vector calculated from the RGB images using equation 1. Then the minimum intensity image is computed over all illumination angles at a given wavelength to filter out the interreflections and subsurface contributions:

$$min \sum_l k_{lj} = I_j^{min}. \qquad (3)$$

The per wavelength difference image is,

$$difference_j = I_j^{max} - I_j^{min}. \qquad (4)$$

This difference image provides information about interreflections around pigment particles as well as multibounce light at layer interfaces: note, for instance, in **Figure 3B** the blue scattered light between bottom layers 1 and 2 as well as the orange scattered light between the top layers 3 and 4. In **Figure 3C** and the corresponding inset detail, the average image is blurry at high magnifications, yet the minimum projection (**Figure 3D**) has resolvable features below 1 µm, as demonstrated by the cluster of three pigment particles marked with a red arrow. This sub-micron resolution is at the theoretical Rayleigh diffraction limit of our imaging system using a 20x, 0.42 NA objective (~0.78 µm at λ = 535 nm). The explanation for this increase in the observed resolution between the average and minimum projections is provided schematically in **Figure 3E**. Two light rays incoming at oblique angles illuminate the sample surface. The forward scattering of the material matrix will produce lobes of volumetrically scattered light which is delineated by dashed lines. In the average projection image, which is the equivalent to standard darkfield illumination (**Figure S2**), patches adjacent to the overlap between the two lobes contain scattered contributions from both directions of illumination which contributes to blur. However, with the minimum projection method, all the light except that confined to the overlap region (gray shaded region, **Figure 3E**) is computationally removed. Thus, in the minimum image, light scattered in the axial and lateral directions is effectively confined to a smaller interaction volume which serves to increase spatial resolution. This observed improvement in resolution over traditional Köhler illumination closely matches the resolution improvements detailed by Ma et al., who describe annular illumination from an array of point sources in transmission.(30)

*Mapping Pigment Distributions with Sparse Modeling.*
While the multiple sample views can be used to extract structural and shape information, the spectral stack contains important information about the artist's process based on the identity and distribution of colorants in the sample which we can calculate. A set of reference reflectance spectra for the pure pigments included in the layered mock-up were formed into a dictionary to perform a sparse linear unmixing and create pigment distribution maps. Our approach has been previously described (41) where we find a sparse representation $w$ of the spectral stack $x$ with respect to a dictionary $D$ composed of known spectral signatures. This is achieved by solving a constrained optimization problem:

$$\text{Minimize } \|w\|_0 \text{ such that } x = Dw \qquad (2)$$

Here, the $l_0$ pseudo-norm counts the number of non-zero elements in $w$. Equation 2 was solved by using a greedy algorithm, orthogonal matching pursuit (OMP), where the number of non-zero elements for each pixel was set to two for this example and optimized sequentially.

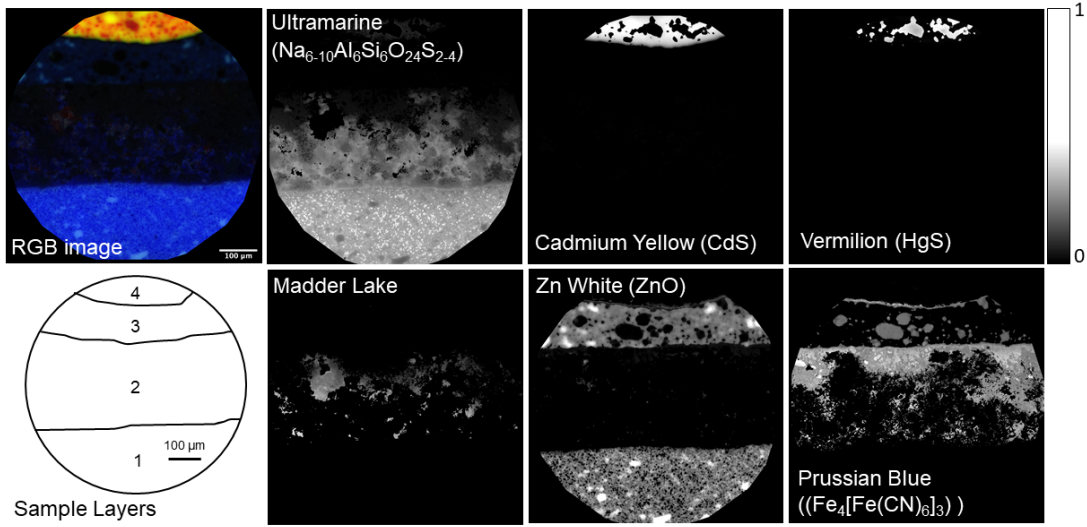

**Figure 4.** Cross-section of the multi-layered mockup sample including the reconstructed RGB image and the corresponding pigment maps created by sparse modeling of the diffuse spectral dataset. The maps were created using an orthogonal matching pursuit (OMP) algorithm with a dictionary comprised of pure pigment spectra (ultramarine, cadmium yellow, vermillion, madder lake, zinc white, and Prussian blue). Distribution weights are normalized between 0 and 1 on a per pixel basis.

The resulting pigment distribution maps for the mockup sample are shown in **Figure 4**. Based on the known content of the layers, the algorithm successfully isolated pigments to the correct strata. The only exception is where the algorithm incorrectly identified Prussian blue in layer 2 of the specimen. This false positive result is not surprising since the dense mixture of madder and ultramarine in this example is highly absorbing and exhibits a spectral shape with an appearance similar to Prussian blue: an almost flat, near zero reflectance profile across the visible range. However, the loading of these high tint pigments in this mockup is extreme, and such a scenario is unlikely to be encountered in a real sample.

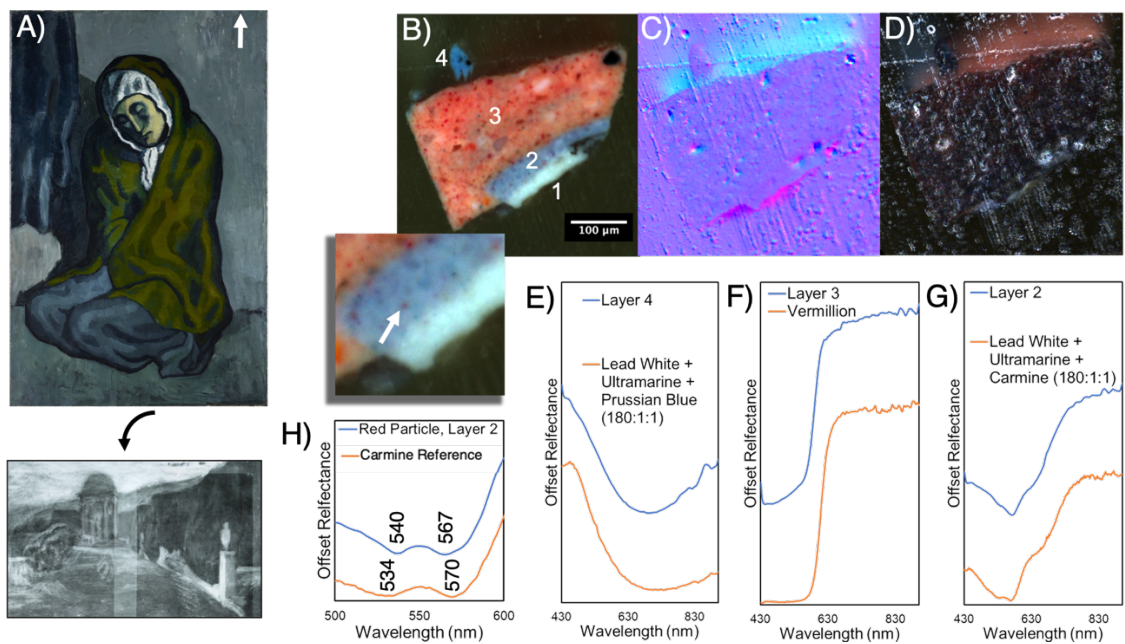

**Figure 5.** A) Photograph of Pablo *Picasso's La Miséreuse accroupie* (1902) and its corresponding X-radiograph rotated by 90° to reveal an underlying landscape painting (image courtesy of the Art Gallery of Ontario). The cross section was removed from the upper right tacking edge, indicated by the white arrow in A), imaged and decomposed into the B) minimum projection, C) surface shape components, and D) the maximum-minimum. E-H) Representative spectra are compared to references from each of the non-white paint layers.

**Case Study Using a Cross-Section from Picasso's *La Miséreuse accroupie*.**

In 1902, the artist Pablo Picasso (1881-1973) painted a crouching woman on top of a landscape scene of a park. This hidden scene was discovered during routine examination by X-radiography (see **Figure 5A,** where the X-radiograph is rotated by 90°) conducted by the Art Gallery of Ontario, where the painting currently resides. To better understand the pigments used to paint the sky of the hidden painting, a small sample was removed from its tacking edge as indicated by the arrow in **Figure 5A**.

The sample was embedded in resin and the surface in **Figure 5B** was revealed by cutting away excess resin using ultramicrotomy. Using the same procedure as described for the mockup cross-section, the spectral microscope was implemented to interrogate the identity and spatial distribution of the pigments distributed in four layers as indicated by the minimum projection image through 10 illumination angles shown in **Figure 5B**. **Figure 5C** provides the surface normal vector map which reveals grooves cut into the sample by the microtomy knife. The image in **Figure 5D** compliments this information and demonstrates surface interreflections from these grooves. In addition, features below the top surface of the transparent resin may be observed as a blurry orange or white glows (corresponding to the top and bottom of the section) that are removed from the minimum image using our filtering procedure.

Average area spectra were extracted from each layer as shown in **Figure 5E-H** and pigment identifications were proposed based on comparison with reference paint mixtures. A summary of the complimentary analytical tests that were performed and the proposed pigment identifications can be found in **Table S1**. In every layer, the microscope allowed for the complete complex heterogeneity of the sample to be examined in detail.

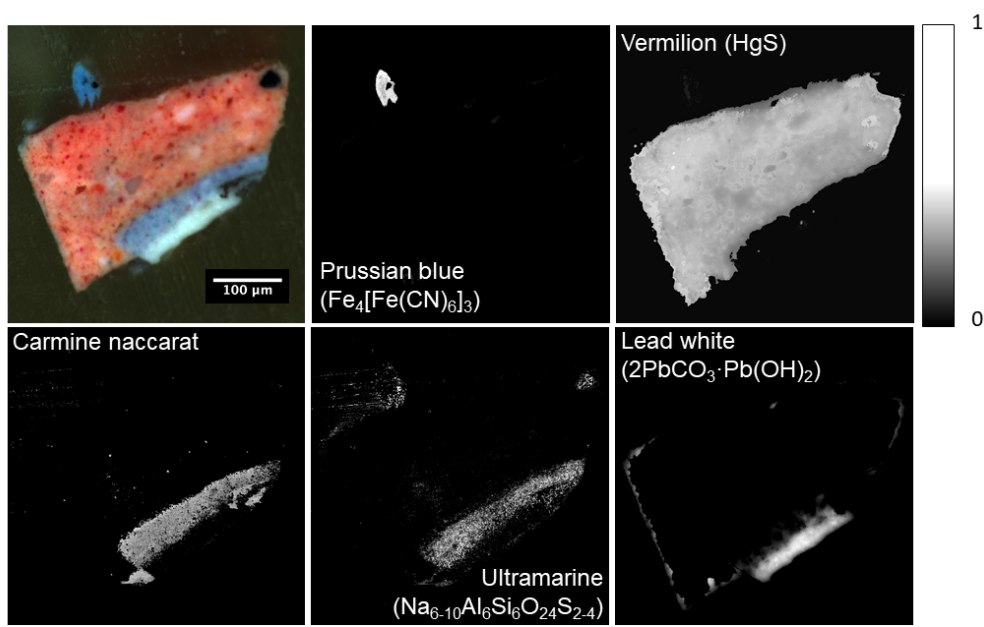

**Figure 6.** Images of a cross section from *La Miséreuse accroupie* (1902): reconstructed RGB image (top left) and the pigment maps for Prussian blue, vermilion, carmine naccarat (an aluminum lake of carminic acid), ultramarine, and lead white. Distribution weights are normalized between 0 and 1.

In the top layer (4) it was proposed, based on the elemental information from SEM-EDS, that there might be more than one blue pigment present. Although many of Picasso's blue period paintings are known to have utilized Prussian blue, the simultaneous use of ultramarine has only been documented in rare instances.(42) A comparison of a reference paint created with a 180:1:1 w/w ratio of lead white : ultramarine : Prussian blue produces an excellent spectral match which was confirmed using Raman spectroscopy (**Figure 5E**). In the thickest subsequent red layer (3), the spectral signature is dominated by the presence of vermilion (**Figure 5F**). However, upon closer inspection there is an anomalous yellow particle as previously highlighted in **Figure 2**. The spectrum of the region (**Figure 2C**) displays a shoulder at 535 nm and matches a cadmium yellow reference (not shown); this was corroborated by the presence of Cd from SEM-EDS, which demonstrates the effectiveness of identifying single particles in a complex matrix with our instrument. In the purple layer (2), the extracted spectrum clearly represents a mixture with multiple absorption bands observed at 540, 567, and approximately 630 nm (**Figure 5G**). SEM-EDS indicated that this layer was aluminum-rich, previously inspiring the hypothesis that this layer contained an organic lake pigment.(31) Consequently, a reference sample containing a 180:1:1 w/w mixture of lead white : ultramarine : carmine naccarat was created. This reference matched the spectral features observed in the historical sample. Indeed, when spectra were extracted from isolated red particles (**Figure 5B, inset**) in both the historical and reference samples, as viewed in **Figure 5H**, characteristic absorption bands can be observed; these are separated by approximately 35-40 nm due to the nature of the ground and excited electronic states of the anthraquinone colorant present.(43) Moreover, recent work on improved protocols for identifying organic reds from reflectance data pushes this identification further to support the specific presence of a cochineal-based rather than plant-based (e.g., madder) pigment and suggest that the substrate used for precipitation was aluminum rather than tin.(44) The unambiguous presence of single cochineal lake pigment particles is clearly indicated from the micro-spectral imaging. The only other technique currently capable of such sensitivity to organic-based colorants is surface-enhanced

Raman spectroscopy which is usually limited to point analysis rather than the mapping demonstrated by our spectral microscope.(45) Lastly, the presence of ultramarine, indicated by the band at 630 nm, was also confirmed by Raman spectroscopy. The sparse modeling of pigment distributions in all the layers is shown in **Figure 6.** In this instance the number of non-zero elements in *w*, from equation 2, was constrained to three. This dataset demonstrates the power of OI spectral microscopy as a sensitive, low-dosage method to characterize pigments in cross-section microsamples that may be challenging to detect with alternative analytical techniques.

**Conclusions**

The OI spectral imaging microscope demonstrated in this work effectively captures the sample heterogeneity observed in complex cultural heritage objects. The spatial distribution of chemical information from reflectance spectra can be routinely captured in addition to surface shape information made possible by the multi-axis movement of the illumination. Also, the hyperspectral datasets are generated using low intensity radiation, which translates to decreased potential for sample damage. Ultimately, as demonstrated from the case study of a cross section sample from *La Miséreuse accroupie*, the OI spectral imaging microscope provides a wealth of molecular data on microscale features and material properties that can be mined computationally.

**Materials and Methods**

*OI Spectral Microscope.* The OI spectral microscope was built on a Cerna modular microscope frame (Thorlabs). A tunable light source consisting of a 300-watt xenon light source filtered through a monochromator (Newport TLS260-300X), enabling down to 5 nm wavelength resolution, was used to illuminate the sample. The broad nature of absorption bands for most pigments in this wavelength range means that the 5 nm resolution is sufficient to capture any unique and identifying spectral features. A 4f arrangement of achromatic lenses (Thorlabs AC254-06/30-A-ML) were used to refocus the beam to a point at a far distance from the focal plane (~20 mm) as is shown in **Figure 1**. Two DT-80 stages (Physik Instrumente) were used to rotate the sample platform and the illumination arm via a microprocessor board (Tiny-G CNC controller) programmed with custom Python scripts. We implemented a scientific complementary metal oxide sensor (sCMOS, Photometrics PrimeBSI) with a 4.2-megapixel array. To optimize data management, most experiments utilized an annular array of azimuthal lighting positions collected at a fixed polar angle of 50° relative to the sample surface normal. To accommodate for the movement of the illumination arm and stage, a long working distance objective (MY20X-804 - 20X Mitutoyo Plan Apochromat Objective, 0.42 NA, 20 mm WD) was used to collect the sample scattering. Spectral image stacks were collected across the visible and near-infrared (450-1000 nm) with an image captured at 2 or 5 nm increments.

*Image Post-Processing.* In a typical experiment, 115 sub images are collected from 430 – 1000 nm before the sample stage is rotated to 9 additional azimuthal angles (every 36°) for a total of 1150 images. The images at each wavelength form a wavelength or lambda stack. For each azimuthal angle, the lambda stacks are registered by the known rotation angle as well as small translations due to micron scale wobble and centering misalignments of the sample platform. The translation transform is determined by a scale-invariant feature transform (SIFT), template matching, or by hand labelling features in each image. The translation is determined on RGB images calculated from each lambda stack: the wavelength images are first collapsed into a XYZ tristimulus image by multiplying each wavelength by a color matching function and summing across all wavelengths to produce X, Y, and Z bands prior to a transformation into RGB color space.(46) After the translation transform is known for each sample rotation angle, it is then applied to all 1150 images to produce a well-aligned four dimensional dataset composed of 2 spatial dimensions plus wavelength and lighting angles. After registration, these multiple angled views of the sample are collapsed into a single image by calculating an average or minimum composite image. This procedure is repeated for each wavelength and the minimum projections are concatenated to create a spectral stack of diffuse-only scattered light. All registration, color transformation, photometric stereo, and matching pursuit steps are implemented as custom written Python scripts in the Fiji image processing suite which are available in the NU-ACCESS Github repositories (https://github.com/NU-ACCESS/).(47)

*Extracting Spectral Information.* To extract spectral information, a white reference target of fused silica (Heraeus, HOD® – High Purity Fused Silica Diffusor, 25 mm diameter x 1 mm thick), was used to normalize the obtained reflectance data between 430 nm and 1000 nm.(48–50) Noise in the measurement increases as the sensitivity of the camera decreases in the NIR range. Fifty dark images obtained with the source shuttered were averaged and then subtracted from both the reflection and white reference images to mitigate the effect of background noise from the sensor. The reflectance for each wavelength at each pixel was calculated using Equation 3, where I is the intensity of the pixel values for the sample images and white reference and D is the average pixel intensity of the set of dark images.

$$Refl_{(\lambda,x,y)} = \frac{\left(I_{image(\lambda,x,y)} - D_{avg(x,y)}\right)}{\left([I_{white(\lambda,x,y)} - D_{avg(x,y)}]_\lambda\right)} \quad (3)$$

*Complementary Analytical Techniques.* SEM images were acquired on a Hitachi S-3400N-II using a 20 kV accelerating voltage and 50 µA probe current. EDS maps were constructed using a higher probe current (70 µA) and a minimum of 30,000 counts per spectrum were required for analysis. Samples were coated in carbon tape to mitigate charge accumulation. Raman spectra were acquired on a Horiba LabRAM HR Evolution Confocal Raman instrument. The excitation wavelength was $\lambda_{ex}$ = 633 nm, using 50x or 100x LWD objectives and a 600 gr/mm grating. Acquisition times ranged from 5-50 s with 3.2 - 5% of the maximum laser power delivered at the sample.

*Painting cross sections.* Reference paintouts were prepared by mixing pigments obtained from Kremer with Galkyd Medium (Gamblin Artist's Colors). Pigment mixtures were based on weight percentages shown in **Table S2** and the alkyd medium was added dropwise until a workable, homogeneous paint mixture was achieved by grinding pigment and medium together on a glass plate. Paint layers were applied evenly to glass slides and dried in ambient conditions or at 40°C. A multi-layered mockup was prepared by painting additional layers with a brush so that each underlying layer was not visible. When dry, samples were excised with a scalpel, mounted as a cross-section in an epoxy resin (Epo-Thin 2, Buehler) and polished using a series of diamond suspensions in water to finish. The sample extracted from *La Miséreuse accroupie* was also mounted in an epoxy resin and an exposed sample surface was achieved by microtomy using a glass knife.

## Acknowledgments

This collaborative initiative is part of the Center for Scientific Studies in the Arts broad portfolio of activities, made possible by the Andrew Mellon foundation. In addition this work made use of the EPIC and Keck II facilities of Northwestern University's NU*ANCE* Center, which has received support from the Soft and Hybrid Nanotechnology Experimental (SHyNE) Resource (NSF ECCS-1542205); the MRSEC program (NSF DMR-1720139) at the Materials Research Center; the International Institute for Nanotechnology (IIN); the Keck Foundation; and the State of Illinois, through the IIN. This work also made use of the MatCI Facility supported by the MRSEC program of the National Science Foundation (DMR-1720139) at the Materials Research Center of Northwestern University. We thank Sandra Webster-Cook and Kenneth Brummel at the Art Gallery of Ontario for the motivation to better understand Picasso's painting materials and methods using a cross-section removed from their painting. The authors also wish to thank Emeline Pouyet for help in preparing the Picasso cross-section through microtomy.